\documentclass{article}
\usepackage{graphicx} 
\usepackage{xurl}
\usepackage{booktabs}
\usepackage{tabularx}
\usepackage{array}
\usepackage{ragged2e}
\usepackage{makecell}
\usepackage[hidelinks]{hyperref}
\hypersetup{
  pdftitle={Feedback Over Form},
  pdfauthor={Charles Junichi McAndrews},
  pdfsubject={Code Generation with Small Language Models},
  pdfkeywords={NEAT, code generation, execution feedback, small language models, HumanEval},
}

\title{Feedback Over Form: Why Execution Feedback Matters More Than Pipeline Topology in 1-3B Code Generation}
\author{Charles Junichi McAndrews}
\date{April 2026}

\begin{document}

\maketitle

\begin{abstract}
    Small language models (1-3B) are practical to run locally, but individually limited on harder code generation tasks. We ask whether composing them into pipelines can recover some of that lost capability. We study code generation pipelines built from 1-3B models with execution feedback, and use a NEAT-inspired evolutionary search to test whether more complex pipeline structure helps beyond a simple refinement loop. We evaluate on HumanEval (164 problems) and sanitized MBPP (427 problems), all with local inference on a single laptop. Self-refinement with execution feedback improves code generation by $>4\sigma$ on both benchmarks. The gains are narrow in mechanism: refinement fixes many runtime errors (especially NameError and SyntaxError), but rarely fixes logic errors such as AssertionError. Within our tested general-purpose model pool, generator identity mattered less than refiner capability: a 1.5B generator paired with a 3B refiner matched a 3B model doing both roles. Early stopping is essential --- without it, every iteration is net-negative. The code-specialized models outperform every general-purpose pipeline configuration, suggesting model specialization matters more than pipeline architecture. Preliminary text-only pipeline experiments without execution feedback did not show gains at this scale. In our constrained search space, evolutionary search mostly rediscovered the same simple generate $\rightarrow$ execute $\rightarrow$ refine loop we found manually, with no clearly significant gain from added topology. Single-evaluation fitness inflates results by 5--7\%, selecting lucky genomes over good ones. On these benchmarks at 1-3B scale, execution feedback mattered more than added pipeline complexity in determining whether composition helped.
\end{abstract}

\section{Introduction}
Small language models (1-3B parameters) are becoming increasingly relevant. These models are small enough for on-device deployment, edge computing, privacy-sensitive applications, and resource-constrained environments. Individually these models are extremely limited and cannot match larger models on complex tasks. This raises a natural question: can composing multiple small models into pipelines exceed what any single small model achieves? This question is well-studied at larger scales (70B+, API Models) where prior work (e.g., MoA, DSPy, AgentConductor) has shown gains from composition. At small scale, this answer is not as clear. Prior work like Self-MoA \cite{selfmoa2025} suggests mixing weak models degrades quality. However, CYCLE \cite{cycle2024} shows self-refinement with execution feedback works. Where exactly is the boundary between composition that helps and composition that hurts at small scale? Our main claim is that execution feedback is the mechanism that makes small-model composition work for code generation. A secondary claim is that, at this scale, more complex topology adds little over a simple generate $\rightarrow$ execute $\rightarrow$ refine loop.

We systematically tested small-model composition on code generation tasks using two standard benchmarks: HumanEval \cite{humaneval2021}, Sanitized MBPP \cite{mbpp2021}. We built pipelines that compose multiple 1-3B models with code execution feedback: a model generates code, an executor runs it, and a refiner attempts to fix errors based on the traceback. We used NEAT \cite{neat2002} to search over pipeline configurations, such as which models fill which roles, how many refinement stages, whether to include an analyzer node, iteration counts, and temperatures. All of these experiments were run locally on a single laptop, demonstrating that meaningful research on model composition is accessible without cloud infrastructure.

We found that execution feedback is the load-bearing mechanism. Self-refinement with execution feedback improved code generation by more than 4 standard deviations on both benchmarks. But the improvement is narrow: refinement reliably fixes runtime errors such as NameError and SyntaxError, while logic errors remain mostly unfixable. The traceback provides a usable signal when the failure is explicit, and much weaker signal when it only reports a wrong answer.

We also found that the generator mattered less than the refiner in the general-purpose model pool we tested. A 1.5B generator paired with a 3B refiner matched a 3B model doing both jobs. Early stopping was essential: without it, every extra refinement pass was net-negative because the refiner often broke code that was already correct.

Our evolutionary search over constrained linear pipelines mostly converged to the same simple generate $\rightarrow$ execute $\rightarrow$ refine loop that manual experimentation also found. One completed run added an analyzer node, but it did not produce a clearly significant gain over the simpler baseline. What mattered more was the capability of the model doing the refinement step. Code-specialized models remained stronger than every general-purpose pipeline configuration we tested.

We also ran exploratory experiments on text-only pipelines, heuristic fitness optimization, and small-model fine-tuning. These mostly produced negative results, but they are secondary to the main code-generation story and we discuss them later as supporting context rather than central evidence. 

\begin{itemize}
    \item We show that execution feedback significantly improves small-model code generation on HumanEval and MBPP, with gains above 4 standard deviations over solo generation.
    \item We provide a mechanism-level analysis showing that refinement mainly fixes runtime errors, while logic errors remain mostly unfixable.
    \item We show that early stopping is load-bearing: without it, every refinement iteration is net-negative because the refiner often breaks already-correct code.
    \item We show that, within the tested general-purpose model pool, refiner capability mattered more than generator identity.
    \item We apply a NEAT-inspired evolutionary search over constrained pipelines and find that it mostly rediscovered simple human-designed topologies rather than yielding a clearly significant improvement over them.
    \item We quantify evaluation noise as a confounder for evolutionary search over stochastic LLM pipelines, showing that single-evaluation fitness inflates results by 5--7
\end{itemize}

\section{Related Work}
\subsection{Self-Refinement with Execution Feedback}
The closest literature to this work is found in CYCLE \cite{cycle2024}. This paper featured same-model self-refinement with execution feedback on code generation benchmarks. CYCLE showed that feeding tracebacks back to the same model improves code generation. To our knowledge, no prior work systematically compares cross-model composition against self-refinement at 1-3B scale. Our work extends CYCLE in two ways: cross-model composition (different generator and refiner) and evolutionary search over pipeline configurations. We also provide mechanistic analysis CYCLE doesn't: Exactly which error types are fixable and why, the early stopping paradox, and regression rates.

\subsection{Pipeline and Workflow Optimization}
AgentConductor \cite{agentconductor2026} uses reinforcement learning (GRPO) to evolve multi-agent topologies with homogeneous workers which required training. Our approach uses NEAT with heterogeneous models and no training. AFlow \cite{aflow2025} and EvoFlow \cite{evoflow2025} were both workflow optimizations but at the scale of 70B+ and API models. We operate at 1-3B with local inference, which is a fundamentally different resource regime. DSPy \cite{dspy2023} is a framework for programmatic prompt optimization for LLM pipelines, but did not search over topology. The key distinction between these works is that they optimized pipelines of large capable models. We're asking whether composition can compensate for individual model weakness at small scale.

\subsection{Model Composition and Ensembles}
Mixture-of-Agents \cite{moa2024} aggregates outputs from multiple LLMs to improve quality. Self-MoA \cite{selfmoa2025} showed that mixing weak models degrades rather than improves quality, which is directionally consistent with our exploratory text-pipeline results. Our experiments suggest a possible boundary at this scale: composition did not show gains on text tasks without verification, but did help on code tasks with execution feedback. FrugalGPT \cite{frugalgpt2023} showed cascade routing to minimize cost while maintaining quality. FrugalGPT instead focused on cost optimization and not capability improvement. Our work differs from these works because we are not routing, ensembling or aggregating. We're composing models into sequential pipelines with execution feedback as the coordination mechanism.

\subsection{Evolutionary Methods for LLMs}
NEAT \cite{neat2002} is the original algorithm for evolving neural networks over augmenting topologies. We adapt it for LLM pipeline topology rather than neural network architecture. EA4LLM \cite{ea4llm2025} surveyed evolutionary approaches for LLM optimization. This paper acknowledges variance in LLM evaluation but doesn't quantify it as a confounder for evolutionary search. We measure this directly showing single-evaluation fitness inflates results by 5-7\%. We are not aware of prior work applying NEAT specifically to code generation pipeline topology. EvoPrompt \cite{evoprompt2023} uses LLM-generated crossover for prompt optimization. We explicitly did not do this, instead hand crafting the prompts for each type of node and having prompt mutation randomly select a prompt from their requisite pool. Controlled Self-Evolution \cite{cse2026} integrated genetic evolution into self-refinement loops for code. They used feedback-guided mutations, but not NEAT-style topology search.

\subsection{Small Model Limitations}
Small Model Learnability Gap \cite{learnabilitygap2025} showed fine-tuning fails at sub-2B scale because format and behavioral objectives interfere. This directly explains our fine-tuning negative result. A general finding in the literature is that small models often lack the capacity to benefit from techniques that work at larger scale (complex prompting, chain-of-thought, multi-agent debate). Our paper contributes to this literature by probing a more specific boundary: in our experiments, small-model composition helped with execution feedback and did not show gains without it. Where prior work documents that small models struggle with complex techniques, we identify execution feedback as the specific mechanism that makes small-model composition viable.

\subsection{Signal-Driven Repair}
A complementary line of work is ARES, which studies adaptive red-teaming and end-to-end repair of policy-reward systems using external reward signals as the driver for iterative correction~\cite{liang2026aresadaptiveredteamingendtoend}. Both ARES and our framework rely on closed-loop correction driven by externally supplied signals, and both find that the quality and specificity of those signals limit how much repair is possible. In ARES, the bottleneck is the reward model: ambiguous or poorly aligned rewards make it difficult to improve the policy. In our setting, explicit traceback signals provide a strong, localized instruction for fixing runtime errors, whereas more ambiguous signals such as bare \texttt{AssertionError} messages offer much weaker guidance. Situating our results alongside ARES clarifies that execution feedback is a particular instance of a broader pattern: signal-driven repair works best when the feedback is explicit, local, and tightly coupled to the underlying error.

\section{Method}
\subsection{Pipeline Architecture}
\begin{figure}[h]
\centering
\includegraphics[width=\textwidth]{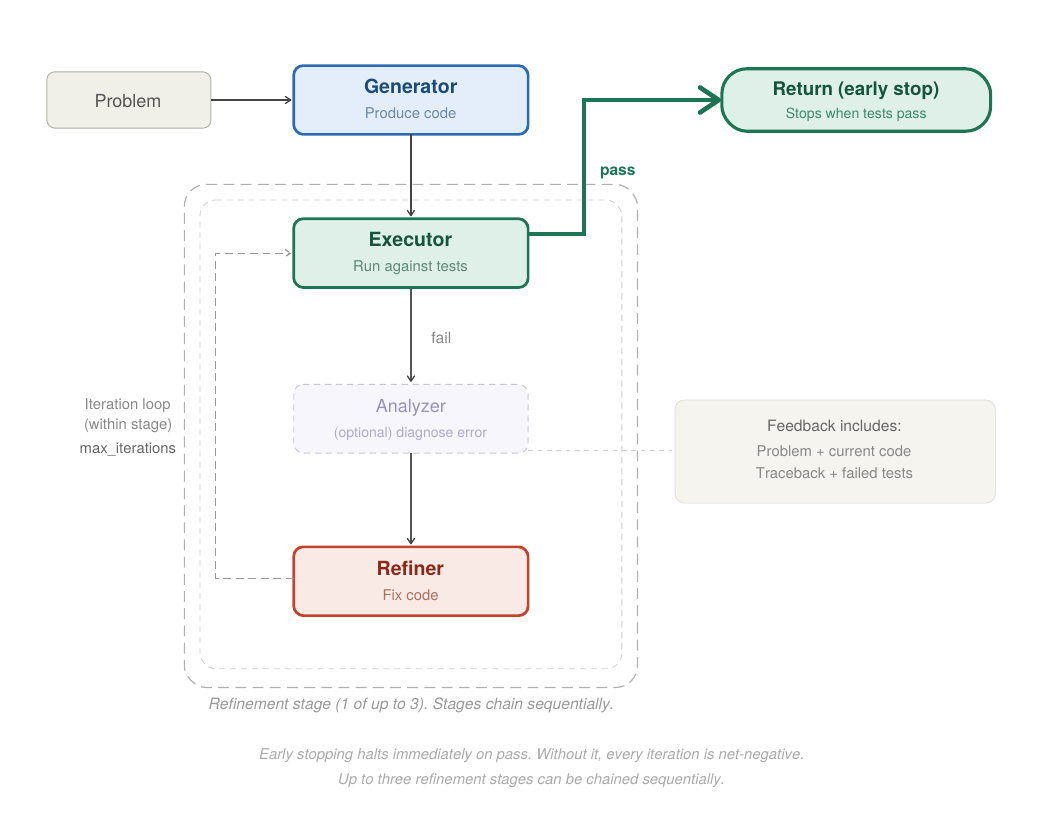}
\caption{Pipeline architecture showing a single refinement stage. The generator produces initial code, the executor runs it against tests, and on failure the optional analyzer and refiner attempt corrections. The loop repeats up to \texttt{max\_iterations} times. Early stopping halts immediately on a passing result. Up to three stages can be chained sequentially.}
\label{fig:pipeline}
\end{figure}

Definitions: A pipeline has one generator followed by 1--3 refinement stages. A stage is one executor $\rightarrow$ optional analyzer $\rightarrow$ refiner block. An iteration is one pass through that stage. Early stopping means the pipeline returns immediately once execution passes.

Node Types:
A pipeline consists of up to 4 node types: Generator, executor, analyzer, and refiner. The generator contains an LLM that receives the problem and produces the requisite code. The generator is made up of the LLM model, the system prompt (1 of 3 possible prompts), and the temperature. The Executor runs input code in a sandboxed subprocess with a 10s timeout. Outputs are pass or fail on the input program. Also outputs the full traceback if the program failed. The analyzer is an optional node. It contains an LLM that receives the failed code and traceback from the executor. The analyzer outputs a natural-language analysis on the error, both the diagnosis on the error and potential corrections. The analyzer has a prompt pool of 2. The refiner contains an LLM that receives the problem, the failed code, the traceback, and the analyzer output if it exists. The refiner has a prompt pool of 3 possible prompts. The model pool used inside the evolutionary search was limited to Gemma3:1B, qwen2.5:1.5B, and llama3.2:3B. We evaluated qwen2.5-coder:3B only as a baseline and did not include it in the search space. These models used default Ollama quantizations (\texttt{Q4\_K\_M}).

Pipeline Architecture:
We used a constrained linear pipeline rather than a free-form DAG. Each pipeline contains one generator followed by 1--3 refinement stages. Each stage contains executor $\rightarrow$ optional analyzer $\rightarrow$ refiner, and each stage can be iterated 1--3 times. We chose this restricted space because more open-ended topologies were unstable in preliminary testing and made search noise worse. \texttt{max\_iterations}=1 means one refinement attempt followed by one execution check. Early stopping was implemented, meaning the pipeline immediately returns the answer if the code passes tests. After a stage exhausts iterations, output feeds to next stage's executor. There is a possible 8 nodes maximum, and a maximum of 3 stages. We deliberately constrained the pipeline to be linear. Free-form topologies failed at this scale in preliminary testing.

Execution Flow:
\begin{enumerate}
    \item The generator produces initial code from the problem description.
    \item The executor runs the code against the benchmark tests.
    \item If execution passes, return immediately.
    \item If execution fails, format the traceback into refinement feedback.
    \item Optionally call the analyzer on the failed code and traceback.
    \item Call the refiner on the problem, failed code, traceback, and analyzer output if present.
    \item Repeat until the stage reaches its iteration limit or the code passes.
    \item If multiple stages are present, feed the latest code into the next stage.
\end{enumerate}

Sandboxing (All within a single executor call):
The python program writes the code and tests to a temp file. This is run via a subprocess, and captured via stdout/stderr/returncode. The error type is parsed from traceback, and the temp file is deleted.

\subsection{NEAT-Inspired Evolutionary Search}
NEAT Adaptation:
We use a NEAT-inspired search over constrained linear pipelines rather than classical free-form NEAT over arbitrary graphs. Our NEAT implementation only operates on the constrained linear pipeline from \S3.1. The "topology" our NEAT implementation searches over is limited to stage count, analyzer presence/absence, which model is assigned to which node, the prompt selected for each node, temperatures, and the iteration counts. Innovation numbers are assigned to refinement stages (not individual connections like classical NEAT). These are used to align stages during crossover so parent genomes can be meaningfully combined.

Seven Mutation Operators (all independent, multiple can fire on the same offspring):
Structural:
\begin{itemize}
    \item add\_refine\_stage (4\%): appends a new executor node and refiner node pair after existing stages. This will be blocked if the pipeline has 3 stages already.
    \item add\_analyzer (5\%): inserts an analyzer node between executor and refiner nodes in a random stage. Blocked if that stage already has an analyzer node.
    \item remove\_node (18\%): 60\% chance  the node removed is a random analyzer, 40\% chance it is the entire last stage that is removed (only if >= 2 stages exist). If no analyzer is present, it will try to remove a stage if there are more than 1 stages. With only 1 stage, it will remove an analyzer node if one exists.
\end{itemize}

Configuration:
\begin{itemize}
    \item swap\_model (25\%): picks a random LLM node and reassigns its model from the model pool.
    \item mutate\_prompt (30\%): picks a random LLM node, selects uniformly from that node type's preset prompt pool. Effective rate is ~15-20\% due to possible re-selection of the current prompt. All prompts are predetermined. No prompts are generated by LLMs during evolution.
    \item adjust\_temperature (20\%): Gaussian jitter $\sigma=0.08$ on a random node's temperature. The value is clipped to [0.05, 1.2].
    \item adjust\_iterations (10\%): +-1 on a random stage's max\_iterations. We biased 60\% towards decrease, with 40\% chance of increasing. Iteration count is clamped to [1, 3].
\end{itemize}

Crossover:
Two parents are selected from previous generation. The offspring will inherit from both. Generator: coin flip, take one parent's generator in its entirety (the full config - model, prompt, temperature). Stages are aligned by the innovation number. If stages match (both parents have same innovation number): coin flip, take one parent's entire stage configuration. For disjoint stages (innovation number only in one parent): More fit parent always passes down their stage. p=0.3 if from a weaker parent. If offspring ends up with more than 3 stages after assembly, drop everything after third stage.

Speciation:
The genomes are grouped into species by compatibility distance. Formula for calculating compatibility distance: stage\_count\_diff * 1.0 (heaviest weighted variable, topology matters the most), analyzer\_count\_diff * 0.5, model\_mismatches * 0.4 each, prompt\_mismatches * 0.2 each, temp\_diff * 0.1 each, iteration\_diff * 0.1 each. The compatibility threshold is adjusted dynamically to maintain 3-5 species. We divide the fitness by its species size to prevent one topology from crowding out alternatives.

Selection and Elitism:
Population size was set to 20. The top 2 are copied and unchanged to next generation (elitism to guarantee the best solutions don't get lost throughout a run). Remaining 18 slots are filled via tournament selection (pick 3 random genomes, take the most fit). Fed into crossover, then fed into mutation. 

Fitness Evaluation:
Each genome was evaluated on a 25-problem HumanEval subset per generation rather than the full 164-problem benchmark. Full evaluation took \textasciitilde15 minutes per genome on the full set, so search used subsets for speed. Our early runs used rotating subsets without stratification; later runs used stratified subsets based on baseline difficulty so fitness was not dominated by easy problems. These subsets were organized into 7 difficulty-balanced columns that rotated across generations in the stratified regime. All genomes within a generation were evaluated on the same 25 problems for fair within-generation comparison. Cross-generation fitnesses are not directly comparable because each generation saw a different subset. Raw fitness was the number of problems solved (pass@1) on that generation's sample. Parsimony penalty was set to \textasciitilde0.02 per node (a 5-node genome loses 0.10 fitness vs. a 3-node genome losing 0.06). As a tiebreaker, we preferred genomes with fewer iterations. Champion fitness was reported via a separate full-benchmark validation pass over all 164 HumanEval problems, not just the 25-problem search subsets. Headline results are 5-run means from that full validation. We ran four searches in total; one early stochastic run was terminated before completion due to complexity growth and is treated as a preliminary failure case rather than a completed experiment. We report two evaluation regimes: stochastic (models sampled at evolved temperatures) and deterministic ($T{=}0$ forced regardless of what evolution selected). The comparison between these two regimes is a key finding in \S4.6. Throughout the paper, we report sigma as a descriptive measure of separation between 5-run means relative to run-to-run variation, using it as an effect-size shorthand rather than a formal hypothesis test.

\subsection{Evaluation Protocol}
Benchmarks:
The first benchmark used was HumanEval \cite{humaneval2021} which included 164 hand-written Python problems. Each problem has a function signature + docstring + 7-10 hidden test assertions. We also used MBPP \cite{mbpp2021} sanitized: 427 crowd-sourced Python problems. Each problem in MBPP has a natural language description and 3 test assertions. Both are standard code generation benchmarks widely used in literature.

Configurations Tested:
\begin{itemize}
    \item llama3.2:3B solo (no pipeline, just generate once)
    \item llama3.2:3B $\rightarrow$ executor $\rightarrow$ llama3.2:3B self-refinement (same model generates and refines, 3 iterations)
    \item qwen2.5:1.5B $\rightarrow$ executor $\rightarrow$ llama3.2:3B cross-model (different generator and refiner nodes, 3 iterations)
    \item NEAT champion: qwen2.5:1.5B generator $\rightarrow$ executor $\rightarrow$ llama3.2:3B analyzer $\rightarrow$ llama3.2:3B refiner (3 iterations)
    \item qwen2.5-coder:3B solo
    \item qwen2.5-coder:3B self-refinement (3 iterations)
\end{itemize}

For consistency with the other pipelines, the validated coder self-refinement baseline in Table~\ref{tab:main-results} used 3 iterations. However, the iteration analysis in \S4.4 showed that 1 refinement step was already sufficient for the coder model, with no meaningful gain from additional iterations.

Validation Protocol:
All reported numbers are 5-run means $\pm$ standard deviation. Each run uses the same benchmark problems under repeated stochastic decoding. We did not fix explicit random seeds across all components, so the run-to-run variance reflects natural sampling noise. We've marked single-run results explicitly wherever they appear. Statistical significance was assessed by computing separation in standard deviations ($\sigma$) between configuration means.

Hardware and Inference:
Everything runs locally on a 128GB M4 Max MacBook Pro. No cloud compute was used for any inference. All models were served via Ollama. Up to 4 models were pinned in memory simultaneously (\texttt{OLLAMA\_MAX\_LOADED\_MODELS=4}). Context length: 4096 tokens (\texttt{OLLAMA\_CONTEXT\_LENGTH=4096}). Total model memory footprint was approximately 8GB for the three-model general-purpose pool. Default Ollama quantizations (\texttt{Q4\_K\_M}) for all models.

\section{Results}
\subsection{Execution Feedback Significance}
We found that self-refinement with execution feedback significantly improved code generation. Both benchmarks show $>4\sigma$ improvement. This is the strongest result in this paper, and the improvement is large enough that we investigated this specific mechanism (i.e. what exactly is the refiner fixing?).

\begin{table}[h]
\centering
\caption{Main results across all configurations and benchmarks. All numbers are 5-run means $\pm$ standard deviation.}
\label{tab:main-results}
\begin{tabularx}{\linewidth}{Xcc}
\hline
\textbf{Configuration} & \textbf{HumanEval (164)} & \textbf{MBPP (427)} \\
\hline
llama3.2:3B solo & 76.6 $\pm$ 3.7 (46.7\%) & 255.8 $\pm$ 5.5 (59.9\%) \\
llama $\rightarrow$ exec $\rightarrow$ llama (self-refine) & 94.0 $\pm$ 2.7 (57.3\%) & 286.2 $\pm$ 2.9 (67.0\%) \\
qwen $\rightarrow$ exec $\rightarrow$ llama (cross-model) & 93.6 $\pm$ 3.0 (57.1\%) & 287.0 $\pm$ 2.0 (67.2\%) \\
NEAT champion & 98.2 $\pm$ 3.4 (59.9\%) & N/A \\
qwen2.5-coder:3B solo & 133.6 $\pm$ 3.4 (81.5\%) & 295.4 $\pm$ 5.4 (69.2\%) \\
qwen2.5-coder:3B self-refine & 139.6 $\pm$ 2.5 (85.1\%) & 307.2 $\pm$ 4.3 (71.9\%) \\
\hline
\end{tabularx}
\end{table}

The NEAT champion was not separately evaluated on MBPP. We did not extend that configuration beyond HumanEval because its advantage over the simpler manual baseline on HumanEval was not statistically significant.

\begin{table}[h]
\centering
\caption{Statistical significance of key comparisons. $\sigma$ denotes separation in standard deviations between configuration means.}
\label{tab:significance}
\begin{tabularx}{\linewidth}{>{\RaggedRight\arraybackslash}Xccc}
\hline
\textbf{Comparison} & \textbf{HumanEval} & \textbf{MBPP} & \textbf{Verdict} \\
\hline
Self-refine vs solo (llama) & +17.4 (4.7$\sigma$) & +30.4 (4.9$\sigma$) & Significant \\
Cross-model vs self-refine & $-$0.4 (0.1$\sigma$) & +0.8 (0.2$\sigma$) & No difference \\
Coder self-refine vs coder solo & +6.0 (1.4$\sigma$) & +11.8 (1.7$\sigma$) & Suggestive \\
Coder self-refine vs best llama & +45.6 (12.4$\sigma$) & +20.2 (4.2$\sigma$) & Significant \\
NEAT champion vs manual best & +4.2 (1.0$\sigma$) & N/A & Not significant \\
\hline
\end{tabularx}
\end{table}

\subsection{Generator vs. Refiner Role}
Across the general-purpose configurations, cross-model refinement matched same-model refinement on both benchmarks. A 1.5B generator paired with a 3B refiner matched a 3B model doing both jobs. This suggests that, in our setup, refiner capability contributed more than generator identity once the generator produced something close enough to repair. However, the generator still needs to produce something close enough for the refiner to fix. See Table~\ref{tab:main-results}.

\subsection{Error Taxonomy}
The tracebacks provided by the executor give unambiguous signal for runtime errors but only symptoms for logical errors. The refinement stage fundamentally fixes runtime errors, and does not provide general debugging on the program. These outcomes were cross-validated on both benchmarks, showing the pattern holds. Because some error types were rare, especially SyntaxError, we report this table as a pattern analysis rather than a precise ranking of fixability.

\begin{figure}[h]
\centering
\includegraphics[width=\textwidth]{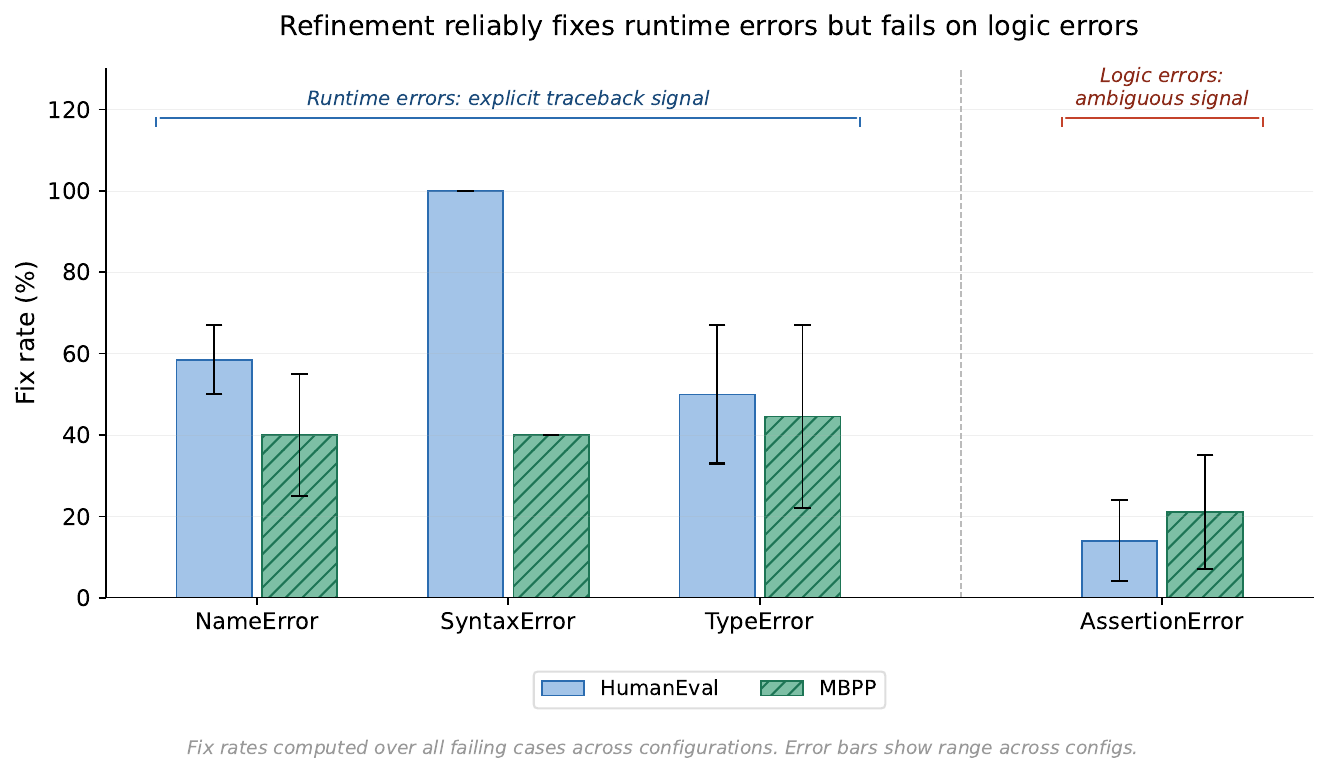}
\caption{Fix rates by error type across both benchmarks. Runtime errors (NameError, SyntaxError) have high fix rates because tracebacks provide explicit signal. Logic errors (AssertionError) have low fix rates because tracebacks only report symptoms. Error bars show the range across configurations.}
\label{fig:error-taxonomy}
\end{figure}

\begin{table}[h]
\centering
\caption{Error taxonomy with fix-rate ranges by error type across both benchmarks; ranges reflect variation across tested configurations, and rare categories should be interpreted cautiously.}
\label{tab:error-taxonomy}
\begin{tabularx}{\linewidth}{>{\RaggedRight\arraybackslash}Xcc>{\RaggedRight\arraybackslash}X}
\hline
\textbf{Error Type} & 
\makecell{\textbf{HumanEval}\\\textbf{Fix Rate}} & 
\makecell{\textbf{MBPP}\\\textbf{Fix Rate}} & 
\textbf{Pattern} \\
\hline
NameError & 50--67\% & 25--55\% & Most fixable \\
SyntaxError & 100\% & 40\% & Fixable (small n) \\
TypeError & 33--67\% & 22--67\% & Middle tier \\
AssertionError & 4--24\% & 7--35\% & Mostly unfixable \\
\hline
\end{tabularx}
\end{table}

\subsection{Early Stopping Prevents Regressions}
Early stopping is load-bearing because refinement is not uniformly helpful. On already-correct problems, the refiner often makes the code worse. On some failing problems, the traceback gives enough signal to recover. This creates an unusual pattern: each additional refinement step is net-negative if forced, but cumulative success still rises when early stopping captures wins and prevents further damage. Tables~\ref{tab:net-iteration}, \ref{tab:early-stopping}, and \ref{tab:regressions} are representative single-run analyses used to illustrate the mechanism; the main benchmark results in Table~\ref{tab:main-results} are 5-run means.

\begin{figure}[h]
\centering
\includegraphics[width=\textwidth]{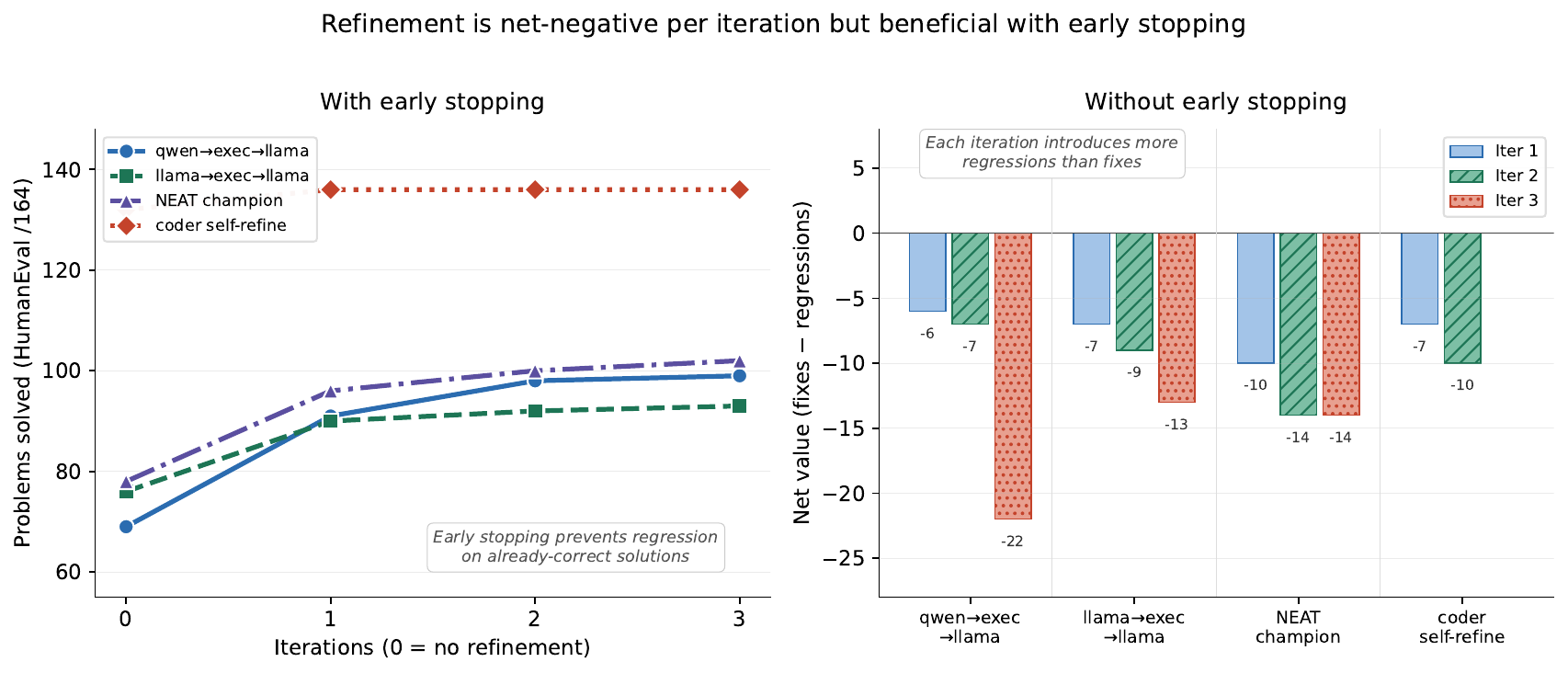}
\caption{The early stopping paradox. Left: cumulative problems solved increases with more refinement iterations when early stopping is used. Right: without early stopping, every iteration is net-negative (more regressions than fixes). Early stopping resolves this by preventing the refiner from touching already-correct code.}
\label{fig:early-stopping}
\end{figure}

\begin{table}[h]
\centering
\caption{Net value per iteration without early stopping (new fixes minus new regressions). Every iteration is net-negative.}
\label{tab:net-iteration}
\begin{tabular}{lccc}
\hline
\textbf{Configuration} & \textbf{Iter 1} & \textbf{Iter 2} & \textbf{Iter 3} \\
\hline
qwen $\rightarrow$ exec $\rightarrow$ llama & $-$6 & $-$7 & $-$22 \\
llama $\rightarrow$ exec $\rightarrow$ llama & $-$7 & $-$9 & $-$13 \\
NEAT champion & $-$10 & $-$14 & $-$14 \\
coder self-refine & $-$7 & $-$10 & 0 \\
\hline
\end{tabular}
\end{table}

\begin{table}[h]
\centering
\caption{Cumulative problems solved with early stopping. Early stopping captures wins and prevents the refiner from touching already-correct code.}
\label{tab:early-stopping}
\begin{tabular}{lcccc}
\hline
\textbf{Configuration} & \textbf{iters=0} & \textbf{iters=1} & \textbf{iters=2} & \textbf{iters=3} \\
\hline
qwen $\rightarrow$ exec $\rightarrow$ llama & 69 & 91 & 98 & 99 \\
llama $\rightarrow$ exec $\rightarrow$ llama & 76 & 90 & 92 & 93 \\
NEAT champion & 78 & 96 & 100 & 102 \\
coder self-refine & 132 & 136 & 136 & 136 \\
\hline
\end{tabular}
\end{table}

\begin{table}[h]
\centering
\caption{Regression analysis: how often the refiner breaks initially-correct code without early stopping.}
\label{tab:regressions}
\begin{tabular}{lccc}
\hline
\textbf{Configuration} & \textbf{Initial Passes} & \textbf{Broke} & \textbf{Break Rate} \\
\hline
qwen $\rightarrow$ exec $\rightarrow$ llama & 69 & 43 & 62\% \\
NEAT champion & 78 & 45 & 58\% \\
llama $\rightarrow$ exec $\rightarrow$ llama & 76 & 33 & 43\% \\
coder self-refine & 132 & 19 & 14\% \\
\hline
\end{tabular}
\end{table}

\subsection{Code-Specialized Models}
Code-specialized models outperformed every general-purpose pipeline we tested on both benchmarks. Adding refinement still helped qwen2.5-coder:3B, but the gain was much smaller than the gap between coder and general-purpose models. Coder's self-refine pipeline against our best general-purpose model pipeline showed significant differences between the two capabilities: qwen2.5-coder:3B pipeline scored +45.6 ($12.3\sigma$) HumanEval and +20.2 MBPP($4.2\sigma$) against the general-purpose model pipeline. This suggests that, in our setting, model specialization mattered more than adding complexity to the tested general-purpose pipelines.

\subsection{NEAT Evolution}
Table~\ref{tab:neat-runs} summarizes the completed evolutionary search runs. Across the three completed runs, search repeatedly converged to simple pipelines. Two champions used the basic generator $\rightarrow$ executor $\rightarrow$ refiner loop, while one added an analyzer node. The analyzer-bearing champion (v3) achieved the strongest validated score, but its advantage over the simpler manual baseline was small: $98.2 \pm 3.4$ versus $94.0 \pm 2.7$, a gap of 4.2 problems or about 1.0 $\sigma$. We therefore treat it as suggestive rather than convincing evidence that added topology helps at this scale.

The completed runs also show that evolutionary search mostly rediscovered human-designed configurations. Run v2 converged to the same llama3.2:3B self-refinement pattern as the manual baseline and matched it almost exactly after 5-run validation: $93.8 \pm 3.3$ versus $94.0 \pm 2.7$. The deterministic search run also converged to the same simple generator $\rightarrow$ executor $\rightarrow$ refiner loop, but with fewer iterations and lower validated performance. Across these runs, evolution confirmed that simple refinement loops are competitive, but did not show a clearly significant advantage for more complex topologies.

The run summaries also reinforce the evaluation-noise story. Among the completed stochastic runs, v3 had the worse search-time score (21/25) but the stronger full-benchmark validation (106/164 single-run; $98.2 \pm 3.4$ across 5 runs). By contrast, v2 looked better during search (22/25) but validated slightly worse. Search-time fitness was therefore useful for within-run selection, but not a reliable proxy for final quality across runs. In Table~\ref{tab:neat-runs}, L3.2-3B denotes llama3.2:3B and Q2.5-1.5B denotes qwen2.5:1.5B.

\begin{table}[t]
\centering
\caption{Completed NEAT-inspired search runs on HumanEval. All 5-run validations used the full 164-problem benchmark.}
\label{tab:neat-runs}
\small
\setlength{\tabcolsep}{3pt}
\begin{tabularx}{\linewidth}{@{}lccXcccl@{}}
\toprule
\textbf{Run} & \textbf{Eval} & \textbf{Strat.} & \textbf{Topology} & \textbf{Itrs} & \textbf{Search best} & \textbf{Full 164} & \textbf{5-run val.} \\
\midrule
v2 & Stoch. & No &
gen(L3.2-3B) \newline $\rightarrow$ exec \newline $\rightarrow$ ref(L3.2-3B)
& 2 & 0.82 & 99/164 & $93.8 \pm 3.3$ \\

v3 & Stoch. & Yes &
gen(Q2.5-1.5B) \newline $\rightarrow$ exec \newline $\rightarrow$ ana(L3.2-3B) \newline $\rightarrow$ ref(L3.2-3B)
& 3 & 0.76 & 106/164 & $98.2 \pm 3.4$ \\

temp0 & Det. & Yes &
gen(Q2.5-1.5B) \newline $\rightarrow$ exec \newline $\rightarrow$ ref(L3.2-3B)
& 1 & 0.66 & 96/164 & $89.8 \pm 5.2^{a}$ \\
\bottomrule
\end{tabularx}

\vspace{1mm}
\footnotesize{$^{a}$ Search used deterministic decoding ($T{=}0$), but 5-run validation used the evolved temperatures. Search-time scores were measured on 25-problem generation samples. L3.2-3B denotes llama3.2:3B and Q2.5-1.5B denotes qwen2.5:1.5B. A preliminary stochastic run was terminated early due to complexity growth and is excluded from the completed runs.}
\end{table}

\subsection{Evaluation Noise}
Single-run evaluation systematically overstated pipeline quality in our experiments. When we re-validated results with 5-run averages, scores dropped by 5--7\% across the main HumanEval configurations. This matters especially for evolutionary search, where noisy single-shot fitness changes which genomes survive. In other words, search can end up selecting lucky genomes instead of genuinely better ones. The same mismatch appears at the run level in Table~\ref{tab:neat-runs}: the strongest validated champion did not come from the run with the highest search-time fitness.

This is a problem specific to evolutionary search: if fitness is evaluated only once per genome, NEAT selects lucky genomes, not good genomes. The 25-problem-per-generation sample size compounded the noise, because one lucky problem is a 4\% point swing. The same pattern appeared at the per-problem level: stronger specialized models were not only better on average, but also more consistent across repeated runs, whereas weaker general-purpose pipelines showed more run-to-run instability on borderline problems. The forward recommendation in future work is to use multi-evaluation fitness (3--5 runs per genome averaged) or deterministic evaluation ($T{=}0$) when using evolutionary methods on stochastic LLM pipelines.

\begin{table}[h]
\centering
\caption{Single-run results vs.\ 5-run validated means on HumanEval. All single-run results were inflated by 5--7\%.}
\label{tab:eval-noise}
\begin{tabular}{lccc}
\hline
\textbf{Configuration} & \textbf{Single-Run} & \textbf{5-Run Mean} & \textbf{Difference} \\
\hline
llama solo & 82 & 76.6 & $-$5.4 \\
qwen $\rightarrow$ exec $\rightarrow$ llama & 101 & 93.6 & $-$7.4 \\
llama $\rightarrow$ exec $\rightarrow$ llama & 99 & 94.0 & $-$5.0 \\
NEAT champion & 106 & 98.2 & $-$7.8 \\
\hline
\end{tabular}
\end{table}

\subsection{Unfixable Problems}
On HumanEval, 7 of 164 problems were never solved by any tested configuration across runs. 5 of these problems were AssertionErrors --- deep logic problems where the 3B models lacked the reasoning capabilities, and we don't attribute this to being a pipeline problem. The other 2 problems are NameErrors where every model in the pool consistently hallucinated the same wrong identifier. The traceback would tell them exactly what was wrong but they produced the same mistake every run. In the MBPP benchmark, all 427 problems were solved by at least one configuration in at least one run. This means that the MBPP benchmark is fully within the 3B model capability. HumanEval's small hard ceiling showed none of our pipeline architectures broke through at this scale, supporting the paper's thesis: pipeline topology isn't the bottleneck, model capability is.

\section{Discussion}
\subsection{Why Execution Feedback Works}
The tracebacks provided an unambiguous signal for runtime errors: "variable Foo not define on line 12" tells the refiner (or analyzer if available) exactly what to fix. However, for the logic errors (AssertionErrors), tracebacks could only provide the symptoms: "expected 5, got 3". The refiner knows that the output is wrong, but is not told explicitly why. This explains the error taxonomy results: high fix rates for NameError/SyntaxError, but low fix rates for AssertionError. Even though the mechanism is narrow, the effect is large. Our results support a simple mechanism: refinement helps when the feedback is explicit, local, and machine-checkable. It helps much less when the signal only says the answer is wrong.

\subsection{Practical Recommendations}
For small-model code generation, use a refinement pipeline when you have an execution oracle such as tests, a compiler, or a validator. In our setup, one refinement step was enough for the coder model, while up to 3 helped weaker general-purpose pipelines. Always use early stopping. Without it, refinement was net-negative. Within the model pool we tested, investing in the refiner mattered more than investing in the generator. Code for the experimental framework and evaluation pipeline is available at \url{https://github.com/L3G/feedback-over-form}.

\subsection{Negative Results: Text Pipelines}
We also ran exploratory text-only pipeline experiments before focusing on code. These experiments did not show evidence that composition helped without external verification, but we do not treat them as a main contribution of this paper because they were not evaluated as systematically as the code results. We include them as supporting context for why execution feedback became the focus.

In a head-to-head comparison, a fan-out pipeline lost decisively to the solo model qwen2.5:3B (1 win, 26 losses, 3 ties under Claude-judged pairwise evaluation with position bias control). This was directionally consistent with Self-MoA \cite{selfmoa2025}, which found that mixing weak models can hurt response quality. We also tried heuristic fitness optimization: the champion optimized for the heuristic scorer went 8 wins, 0 ties, and 22 losses on the heuristic, but only 5 wins, 13 ties, and 12 losses when judged by Claude. This is a textbook instance of Goodhart's Law, where the fitness proxy diverged from the quality measure we actually cared about.

These text-pipeline failures directly motivated the pivot to execution feedback, as both our exploratory results and prior literature suggested that composition was much more likely to help when paired with an external verification signal. In that sense, the pipeline structure itself mattered less than whether the system had access to reliable feedback.

\subsection{Negative Results: Fine-Tuning at Small Scale}
We also ran exploratory fine-tuning experiments to see whether the pipeline behavior could be distilled into a smaller model. All methods we tried collapsed on behavioral tasks at sub-2B scale. This is consistent with the Small Model Learnability Gap \cite{learnabilitygap2025} which found that at sub-2B, format and personality are separable concerns whose joint training causes collapse. This negative result scoped our contribution as a framework that operates at inference time through composition, and not through training.

\subsection{NEAT's Role and Limitations}
As shown in our results, NEAT confirmed the manual baselines rather than exceeding them. A more expressive search space, larger population, or stronger model pool might still change that result, so we treat this as a negative finding under our search regime rather than a universal statement about topology search. However, NEAT's repeated convergence to simple pipelines suggests that the manual design is competitive in this search space, though we do not take that as proof that it is globally or even locally optimal. The search space was intentionally constrained to linear pipelines, 3 stages maximum, and 3 models. However, a richer search space may yield different results. We added parsimony pressure (-0.02 fitness per node) to prevent complexity accumulation. This was a hand-tuned hyperparameter to bias towards simplicity. The population size of 20 is also small by NEAT standards. With larger populations, one would be able to explore more diverse topologies before converging. One open question: Is topology not the bottleneck at this scale, or did our search not look hard enough?

\subsection{Limitations}
We only tested against two benchmarks: HumanEval and MBPP (sanitized). Both are relatively easy and well-studied. The strongest baseline in the paper, qwen2.5-coder:3B, was evaluated separately and was not included in the evolutionary search space. The only programming language used was Python. All of the models used were between 1-3B in scale. Results may differ at different model sizes, such as 7B+ where models may have more capability to leverage complex topologies. The framework required an execution oracle, which limits the applicability of this strategy to domains with automated verification. Parsimony penalty was hand-tuned, so different parsimony values would produce different NEAT outcomes. The population size of 20 is small, so larger populations could find better topologies. There were no explicit random seeds in validation. The variance came from natural sampling, not controlled sampling. We did not test on harder benchmarks, where the capability ceiling might create more room for pipeline improvements via wider spread of fitness. The 5-run validation catches gross noise, but 5 runs is still a small sample for precise confidence intervals.

\section{Conclusion}
We tested small-model composition for code generation at 1-3B scale and found that execution feedback is the mechanism that makes refinement work. Across HumanEval and MBPP, self-refinement with execution feedback improved over solo generation by more than 4 standard deviations. The gains were narrow in mechanism: refinement mainly fixed runtime errors, depended on early stopping to avoid regressions, and depended more on refiner capability than generator identity within the model pool we tested. Our evolutionary search mostly rediscovered the same simple refinement loop as the manual baseline rather than showing a clearly significant gain from added topology.

These results are specific to 1-3B scale, Python, and code tasks with test suites. Whether these findings hold at 7B+ where models have more capacity to leverage complex topologies is an open question. Harder benchmarks (LiveCodeBench, SWE-bench) with lower baseline performance may create more room for pipeline improvements. The broader pattern may expand to other domains with automated verification, such as SQL execution, compiler feedback, or proof checking. But this paper only tests Python code generation, so any claim beyond that should be treated as a hypothesis rather than a result.

\bibliographystyle{plain}
\bibliography{references}

@inproceedings{cycle2024,
  title={CYCLE: Learning to Self-Refine the Code Generation},
  booktitle={Proceedings of the Association for Computing Machinery on Object-Oriented Programming, Systems, Languages, and Applications (OOPSLA)},
  year={2024},
  url={https://2024.splashcon.org/details/splash-2024-oopsla/15/CYCLE-Learning-to-Self-Refine-the-Code-Generation}
}

@article{agentconductor2026,
  title={AgentConductor: Topology Evolution for Multi-Agent Competition-Level Code Generation},
  author={Wang, Siyu and others},
  journal={arXiv preprint arXiv:2602.17100},
  year={2026}
}

@article{aflow2025,
  title={AFlow: Automating Agentic Workflow Generation},
  author={{Anonymous}},
  journal={International Conference on Learning Representations (ICLR)},
  year={2025},
  url={https://arxiv.org/abs/2410.10762}
}

@article{evoflow2025,
  title={EvoFlow: Evolving Diverse Agentic Workflows On The Fly},
  author={{Anonymous}},
  journal={arXiv preprint arXiv:2502.07373},
  year={2025}
}

@article{dspy2023,
  title={DSPy: Compiling Declarative Language Model Calls into Self-Improving Pipelines},
  author={Khattab, Omar and others},
  journal={arXiv preprint arXiv:2310.03714},
  year={2023}
}

@inproceedings{moa2024,
  title={Mixture-of-Agents Enhances Large Language Model Capabilities},
  author={Wang, Junlin and Wang, Jue and Athiwaratkun, Ben and Zhang, Ce and Zou, James},
  year={2024},
  url={https://arxiv.org/abs/2406.04692}
}

@article{selfmoa2025,
  title={Is Mixing Different Large Language Models Beneficial?},
  author={Li, Wenzhe and Zhang, Zhibin and Zhang, Zhibin and others},
  journal={arXiv preprint arXiv:2502.00674},
  year={2025},
  note={Introduces Self-MoA: outperforms MoA by 6.6\% on AlpacaEval 2.0 and 3.8\% avg. on MMLU/CRUX/MATH}
}

@article{frugalgpt2023,
  title={FrugalGPT: How to Use Large Language Models While Reducing Cost and Improving Performance},
  author={Chen, Lingjiao and Zaharia, Matei and Zou, James},
  journal={arXiv preprint arXiv:2305.05176},
  year={2023}
}

@article{neat2002,
  title={Evolving Neural Networks through Augmenting Topologies},
  author={Stanley, Kenneth O and Miikkulainen, Risto},
  journal={Evolutionary Computation},
  volume={10},
  number={2},
  pages={99--127},
  year={2002},
  publisher={MIT Press}
}

@article{ea4llm2025,
  title={EA4LLM: A Gradient-Free Approach to Large Language Model Optimization via Evolutionary Algorithms},
  author={Liu, WenTao and Song, Siyu and Hao, Hao and Zhou, Aimin},
  journal={arXiv preprint arXiv:2510.10603},
  year={2025}
}

@article{evoprompt2023,
  title={EvoPrompt: Connecting LLMs with Evolutionary Algorithms Yields Powerful Prompt Optimizers},
  author={Zhong, Weixiao and Cui, Liang and Liang, Shu and Zhang, Shu and Li, Chenyang and Liu, Yue and Miao, Xing and Wang, Shuming and Liu, Qiang},
  journal={arXiv preprint arXiv:2309.08532},
  year={2023}
}

@article{cse2026,
  title={Controlled Self-Evolution for Algorithmic Code Optimization},
  author={{Anonymous}},
  journal={arXiv preprint arXiv:2601.07348},
  year={2026}
}

@inproceedings{learnabilitygap2025,
  title={Small Models Struggle to Learn from Strong Reasoners},
  author={Li, Yuetai and Yue, Xiang and Xu, Zhangchen and Jiang, Fengqing and Niu, Luyao and Lin, Bill Yuchen and Ramasubramanian, Bhaskar},
  booktitle={Findings of the Association for Computational Linguistics: ACL 2025},
  year={2025},
  url={https://aclanthology.org/2025.findings-acl.1301}
}

@article{humaneval2021,
  title={Evaluating Large Language Models Trained on Code},
  author={Chen, Mark and Tworek, Jerry and Jun, Heewoo and Yuan, Qiming and Pinto, Henrique Ponde de Oliveira and Kaplan, Jared and Edwards, Harri and Burda, Yuri and Joseph, Nicholas and Brockman, Greg and others},
  journal={arXiv preprint arXiv:2107.03374},
  year={2021}
}

@article{mbpp2021,
  title={Program Synthesis with Large Language Models},
  author={Austin, Jacob and Odena, Augustus and Nye, Maxwell and Bosma, Maarten and Michalewski, Henryk and Dohan, David and Jiang, Ellen and Cai, Carrie and Terry, Michael and Le, Quoc and others},
  journal={arXiv preprint arXiv:2108.07732},
  year={2021}
}

@misc{liang2026aresadaptiveredteamingendtoend,
  title        = {ARES: Adaptive Red-Teaming and End-to-End Repair of Policy-Reward System},
  author       = {Jiacheng Liang and Yao Ma and Tharindu Kumarage and Satyapriya Krishna and Rahul Gupta and Kai-Wei Chang and Aram Galstyan and Charith Peris},
  year         = {2026},
  eprint       = {2604.18789},
  archivePrefix= {arXiv},
  primaryClass = {cs.AI},
  url          = {https://arxiv.org/abs/2604.18789},
}

\end{document}